\documentclass[prb,twocolumn,showpacs,amssymb,amsmath]{revtex4}

\usepackage{graphicx}
\usepackage{bm}
\def\be{\begin{equation}}
\def\ee{\end{equation}}
\def\bea{\begin{eqnarray}}
\def\eea{\end{eqnarray}}
\newcommand{\matr}[4]{{\left(\begin{array}{cc} #1&#2\\#3&#4\\\end{array}\right)}}
\newcommand{\vect}[2]{{\left(\begin{array}{c} #1\\#2\\\end{array}\right)}}
\renewcommand{\vec}{\mathbf}

\renewcommand{\vr}{\vec{r}}

\newcommand{\ve}{\vec{e}}
\newcommand{\vsigma}{\mbox{\boldmath $\sigma$}}

\newcommand{\vA}{\vec{A}}

\newcommand{\vp}{\vec{p}}

\begin{document}
\title{Interplay of Aharonov-Bohm and Berry phases in gate-defined graphene quantum dots}

\author{Julia Heinl}
\author{Martin Schneider}
\author{Piet W. Brouwer}
\affiliation{
Dahlem Center for Complex Quantum Systems and Institut f\"ur Theoretische Physik,
Freie Universit\"at Berlin, Arnimallee 14, 14195 Berlin, Germany
}
\date{\today}

\begin{abstract}
We study the influence of a magnetic flux tube on the possibility to electrostatically confine electrons in a graphene quantum dot. Without magnetic flux tube, the graphene pseudospin is responsible for a quantization of the total angular momentum to half-integer values.
On the other hand, with a flux tube containing half a flux quantum, the Aharonov-Bohm phase and Berry phase precisely cancel, and we find a state at zero angular momentum that cannot be confined electrostatically. In this case, true bound states only exist in regular geometries for which states without zero-angular-momentum component exist, while non-integrable geometries lack confinement. We support these arguments with a calculation of the two-terminal conductance of a gate-defined graphene quantum dot, which shows resonances for a disc-shaped geometry and for a stadium-shaped geometry without flux tube, but no resonances for a stadium-shaped quantum dot with a $\pi$-flux tube.
\end{abstract}

\maketitle

\section{Introduction}
In recent years, graphene has emerged as a promising material for future nanoelectronical devices.\cite{CastroNeto,Beenakker,Geim,DasSarma} The possibility to confine electrons is of particular relevance in this context. Experimental activity concentrates on confinement in quantum dots realized with etched graphene structures\cite{Ponomarenko,Stampfer,Jacobsen} or graphene nanoflakes.\cite{Hamalainen} Electrostatic confinement with the help of metal gates, which is standard in semiconductor heterostructures, is problematic due to the absence of a band gap in monolayer graphene. In particular, an electron that approaches a region of graphene with zero carrier density --- the closest approximation to an ``electrostatic barrier'' in graphene --- will penetrate this region with unit probability if at normal incidence. This phenomenon is known as ``Klein tunneling''.\cite{Klein,Katsnelson,CheianovFalko} Theoretical proposals suggest to use magnetic instead of electric fields to shape quantum dots\cite{DeMartino} or 
induce a gap in the spectrum.\cite{Pal}

\begin{figure}[t]
\includegraphics[width=2.0in]{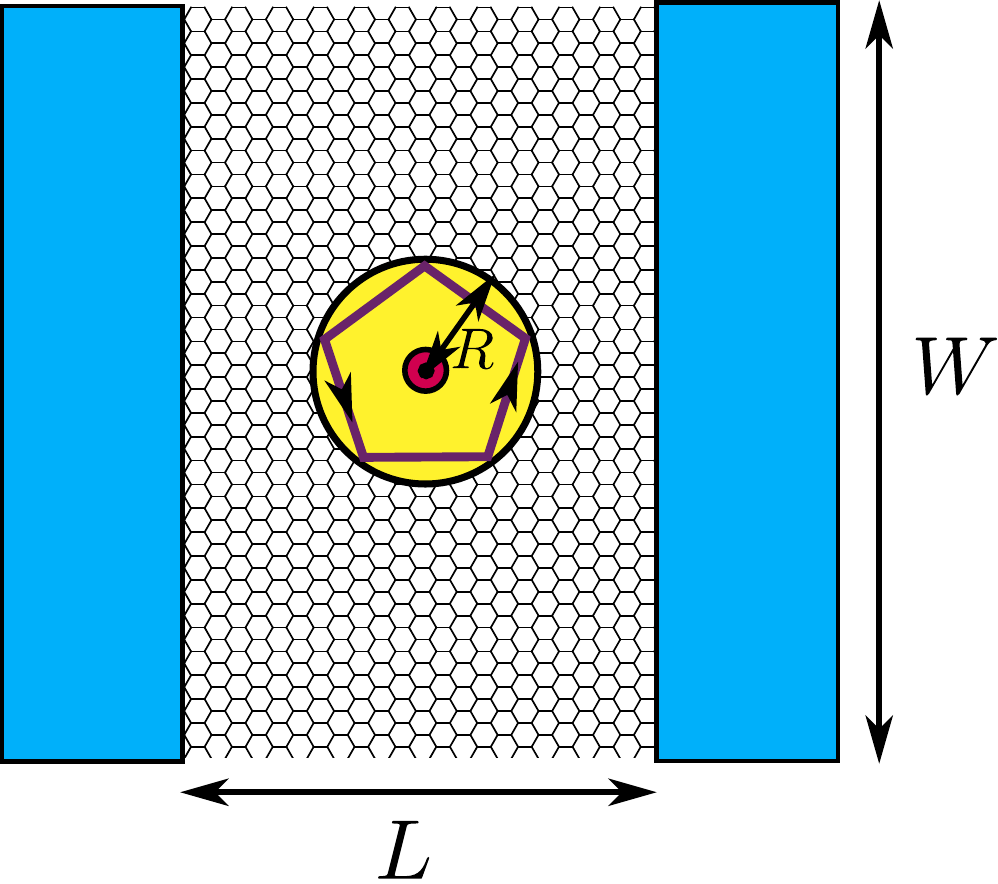}
\caption{(Color online) The geometry under consideration: A quantum dot (here with circular shape), consisting of a region of graphene with a nonzero spatially uniform carrier density surrounded by a carrier-free graphene layer, which is coupled to leads in a two-terminal geometry. The total system has rectangular shape of dimension $L\times W$; the size of the quantum dot is denoted $R$. In this article we consider the effect of a magnetic flux tube through the quantum dot (indicated in red). If the flux tube carries half a flux quantum, the Aharonov-Bohm phase precisely cancels the Berry phase that is accumulated in a cyclic orbit inside the quantum dot.}
\label{fig:geometry}
\end{figure}

The statement that one cannot confine electrons in graphene using gate potentials can be circumvented in certain special cases.\cite{Bardarson09} The reason is that Klein tunneling is effective only at perpendicular incidence, while the reflection probability sharply increases away from normal incidence. Certain integrable geometries, such as a disc,\cite{Pal,calvo2011} allow states that exclude perpendicular incidence, so that electrons can be effectively confined in a disc-shaped region of graphene with finite carrier density, surrounded by a carrier-free ({\em i.e.,} undoped, intrinsic) graphene sheet. On the other hand, for geometries with a chaotic classical dynamics, no such exclusion of perpendicular incidence is possible, and one may expect that no bound states exist in this case. In Ref.\ \onlinecite{Bardarson09}, as well as in later studies,\cite{TitovOstrovsky,Schneider11,Pal} a circular and a stadium-shaped quantum dot, as prototypes of integrable and chaotic geometries, were embedded in a 
carrier-free graphene region and coupled to source and drain 
contacts, as shown schematically in Fig.\ \ref{fig:geometry}. Bound states are then revealed as sharp resonances in the two-terminal conductance.

Interestingly, the conductance of a carrier-free graphene sheet with a stadium-shaped and disc-shaped quantum dot showed resonant features that were {\em quantitatively} different, but {\em qualitatively} similar.\cite{Bardarson09,Schneider11} The quantitative difference concerns the scaling of the resonance widths with the coupling to the leads, which is determined by the ratio $R/L$ of the quantum dot size $R$ and the distance $L$ between the source and drain contacts, see Fig.\ \ref{fig:geometry}. Whereas for the stadium-shaped quantum dot the width was proportional to $R/L$ for all resonances, the disc-shaped quantum dot also featured much narrower resonances, with a width that scaled proportional to $(R/L)^n$ with $n \ge 3$. The qualitative similarity was that in the limit $R/L \to 0$ both systems showed conductance resonances at all. This contradicts the naive classical expectation that there should be no resonances for a chaotic geometry, because in a chaotic geometry each electron eventually hits the 
dot boundary at perpendicular incidence, and an electron that hits the dot boundary at perpendicular incidence exists the quantum dot with unit probability. No resonant structures should exist if the escape probability is unity after a finite time.

This deviation from the naive classical expectation can be attributed to the Berry phase in graphene. In graphene, electrons are assigned a pseudospin that corresponds to the sublattice degree of freedom. The pseudospin is locked to momentum. Upon completion of a full rotation the electron collects a Berry phase of $\pi$. This Berry phase has the important consequence that the lowest possible angular momentum is $\frac{\hbar}{2}$. Perpendicular incidence on the boundary of the quantum dot corresponds to zero angular momentum, so that no states with perpendicular incidence on the surface exist. This then explains why a stadium-shaped quantum dot still shows conductance resonances, in spite of the naive classical expectation that geometries with a chaotic classical dynamics can not be used to confine electrons.

In order to support these arguments, in this article we study gate-defined graphene quantum dots in which the Berry phase is compensated by the Aharonov-Bohm phase from a magnetic flux tube through the quantum dot. (The interplay of the two phase shifts is also used experimentally to identify Berry phase effects, see, {\em e.g.}, Refs.\ \onlinecite{yau2002,qu2011}.) If the magnetic flux tube carries half a flux quantum (``$\pi$ flux''), the Aharonov-Bohm phase and the Berry phase collected along a closed trajectory around the flux tube precisely cancel. We find that with a $\pi$ flux the system can reach a state with zero kinematic angular momentum, that cannot be confined by means of gate potentials. 

The $\pi$-flux tube has qualitatively different consequences for disc-shaped and stadium-shaped geometries. For the disc-shaped geometry, states have a well-defined kinematic angular momentum. While the states with zero kinematic angular momentum are no longer confined, states with nonzero kinematic angular momentum remain confined to the quantum dot. Hence, for the disc-shaped quantum dot the inclusion of the $\pi$-flux tube eliminates some of the resonances, but not all. On the other hand, for the stadium-shaped geometry, all states have a component in the zero-angular-momentum channel, so that inclusion of the $\pi$-flux tube leads to the suppression of all resonances.\cite{Silvestrov}

The remaining part of the paper is organized as follows: In Section \ref{sec:boundstates} we calculate the bound states of a disc-shaped quantum dot in the presence of a magnetic flux tube. We find, that the asymptotic behavior of the zero-angular-momentum state is the same as for a free circular wave. Hence, no bound state can exist in this channel. In Section \ref{sec:twoterminal} we present a numerical calculation of the two-terminal conductance setup of Fig. \ref{fig:geometry}, for a circular and a stadium-shaped quantum dot. Upon inclusion of the $\pi$-flux tube, we find that sharp resonances persist for the circular dot, while the conductance becomes featureless for the stadium dot in the limit $R/L \to 0$. We conclude in Section \ref{sec:conclusion}. 

\section{Disc-shaped quantum dot}
\label{sec:boundstates}

The electrostatically-defined graphene quantum dot is described by the Hamiltonian
\begin{equation}
 \label{eq:H0}
  H_0=v_{\rm F} (\vp+e\vA) \cdot \vsigma + V(\vr),
\end{equation}
where $v_{\rm F}$ is the Fermi velocity $v_{\rm F}$ and $\vsigma=(\sigma_x,\sigma_y)$ are the Pauli matrices. The gate potential $V(\vr)$ is nonzero and constant inside the quantum dot, and zero elsewhere,
\begin{equation}
 V(\vr)=\begin{cases}
         -\hbar v_{\rm F} V_0, & r<R\\
         0,  & r>R,
        \end{cases}
\end{equation}
where $R$ is the radius of the disc-shaped dot. The constant $V_0$ has the dimension of inverse length. We choose $V_0 > 0$, so that the quantum dot is electron doped. The potential $V(\vr)$ is smooth on the scale of the lattice constant, justifying our description in terms of a single Dirac point. The choice of a spatially uniform potential inside dot makes a closed-form solution of the wavefunctions possible and allows for a straightforward comparison to the classical dynamics in the quantum dot, but it is not essential for the existence of bound states.\cite{downing2011,mkhitaryan2012} The structure of quasibound states in the inverted setup (zero potential inside, nonzero outside) was considered in Refs.\ \onlinecite{matulis2008,hewageegana2008}. 

The vector potential corresponding to the magnetic flux line is
\begin{equation}
  \vA(\vr)=\frac{h}{e} \frac{\Phi}{2\pi r} \hat{\vec{e}}_{\theta},
\end{equation}
where $\hat{\vec{e}}_{\theta}$ is the unit vector for the azimuthal angle, and $\Phi$ is the magnetic flux measured in units of the flux quantum $h/e$. 
In polar coordinates, the kinetic part of the Hamiltonian then reads
\begin{equation}
  v_{\rm F} (\vp+e\vA) \cdot \vsigma=-i \hbar v_{\rm F}\matr{0}{D_-}{D_+}{0},
\end{equation}
where we defined the operators
\begin{equation}
 D_{\pm}=e^{\pm i \theta} \left(\partial_r \pm \frac{i}{r} \partial_{\theta} \mp \frac{\Phi}{r} \right).
\end{equation}

With our choice of the vector potential, the Hamiltonian is invariant under rotation, hence we can look for eigenstates of the total angular momentum $j_z=l_z+\frac{\hbar}{2}\sigma_z$. They have the form
\begin{equation}
 \label{waveansatz}
 \psi_m(\vr)=e^{im\theta}\vect{e^{-i\frac{\theta}{2}}\varphi_{m,+}(r)}{e^{i\frac{\theta}{2}}\varphi_{m,-}(r)},
\end{equation}
where $m = \pm 1/2$, $\pm 3/2$, \ldots. Inside the dot, for $r < R$, the radial wave functions $\varphi_{m,\pm}$ are determined by the coupled equations
\begin{align}
  \left(\partial_r-(m-\tfrac{1}{2})\tfrac{1}{r}-\tfrac{1}{r}\Phi\right)\varphi_{m,+}(r)=iV_0\varphi_{m,-}(r),\nonumber\\
 \left(\partial_r+(m+\tfrac{1}{2})\tfrac{1}{r}+\tfrac{1}{r}\Phi\right)\varphi_{m,-}(r)=iV_0\varphi_{m,+}(r). 
\end{align}
Outside the dot the equations decouple, and the radial wave functions show a power law behavior
\begin{equation}
 \varphi_{m,+}(r)=a_+ r^{m-1/2+\Phi}, \quad \varphi_{m,-}(r)=a_- r^{-m-1/2-\Phi},
\end{equation}
with coefficients $a_{\pm}$.

\subsection{Without flux tube}

We first review the solutions without flux tube, for $\Phi=0$.\cite{Bardarson09} With the requirement that the wavefunction is regular for $r\rightarrow 0$, we find for the solution inside the dot
\begin{align}
 \varphi_{m,+}(r)&= J_{|m-1/2|}(V_0 r), \nonumber\\
 \varphi_{m,-}(r)&=i  \mathrm{sgn}(m) J_{|m+1/2|}(V_0 r),
\end{align}
where $J_{n}(x)$ is the Bessel function.
Outside the dot, the wave function must not diverge, which gives the constraints $a_+=0$ ($m>0$) and $a_-=0$ ($m<0$).
From continuity of the wavefunction at $r=R$, we find the resonance condition
\begin{equation}
 J_{|m|-1/2}(V_0 R)=0.
\end{equation}
The wavefunction outside the dot is decaying as $\propto r^{-(|m|+1/2)}$. 

In Section \ref{sec:twoterminal} we connect the quantum dots and the surrounding undoped graphene layer to source and drain contacts. The distance between the contacts is denoted $L$ and the quantum dot is placed halfway between the contacts, see Fig.\ \ref{fig:geometry}. In the limit $L \gg R$, the bound states are then revealed as resonances in the two-terminal conductance as a function of the gate potential $V_0$. These resonances have a finite width $\Gamma$, which can be estimated as\cite{Bardarson09} $\Gamma R \sim {|\psi(L)|^2 L}/{|\psi(R)|^2 R}$. We conclude, that the width of the resonances without flux tube scales as 
\begin{equation}
  \Gamma R \propto \left( \frac{R}{L} \right)^{2|m|}. \label{eq:GammaNoFlux}
\end{equation}
For $|m| = 1/2$ the wavefunction decays proportional to $1/r$, which is marginally non-normalizable. Despite the absence of a bound state in the strict sense, the conductance nevertheless shows a resonance, with a width $\Gamma R \propto (R/L)$.\cite{Bardarson09,TitovOstrovsky,Schneider11}

\subsection{With flux tube}

We now consider a disc-shaped quantum dot with a flux tube carrying half a flux quantum ($\Phi=1/2$) --- a ``$\pi$ flux'' --- at its center. The results take a form similar to those without flux tube if we consider the kinematical orbital angular momentum,
\begin{equation}
  l_{z,{\rm kin}}=[\vr\times(\vp+e\vA)]_z,
\end{equation}
instead of the canonical angular momentum. With the inclusion of a $\pi$-flux, we then find $l_{z,{\rm kin}}=l_z+\frac{\hbar}{2}$. The wavefunctions from Eq. \eqref{waveansatz} are then eigenstates of $j_{z,{\rm kin}}$ with eigenvalue $\mu \hbar$, where $\mu=m+1/2$, i.e. the kinematical angular momentum takes on {\it integer} values. For $\mu\neq 0$ the calculation for the bound states proceeds in the same way as without flux, and we find that the resonance condition is given by 
\begin{equation}
  J_{|\mu|-1/2}(V_0 R)=0.
\end{equation}
Outside the dot, the wavefunction decays proportional to $r^{-(|\mu|+1/2)}$. We conclude that, if the dot and the surrounding undoped graphene layer are contacted to source and drain reservoirs, the width $\Gamma$ of the resonances in the two-terminal conductance scales as
\begin{equation}
  \Gamma R \propto \left(\frac{R}{L}\right)^{2|\mu|}. \label{eq:GammaFlux}
\end{equation}

The state with zero kinematical angular momentum ($\mu=0$) however is special: First of all, inside the dot, the wavefunction is of the form
\begin{equation}
 \label{eq:mu0}
 \psi(\vr)=b_1\vect{e^{-i \theta} J_{1/2}(V_0 r)}{i Y_{1/2}(V_0 r)}+b_2 \vect{e^{-i \theta} Y_{1/2}(V_0 r)}{-i J_{1/2}(V_0 r)}.
\end{equation}
Recalling that the half-integer Bessel functions take the simple form $J_{1/2}(x)= \sqrt{2/\pi x} \sin x$, and $Y_{1/2}(x)=-\sqrt{2/\pi x} \cos x$, we see that $\psi(\vr)$ diverges as $1/\sqrt{V_0 r}$ at the origin, and that there is no non-trivial choice of coefficients $b_1$ and $b_2$ which removes this divergence. The root of this singular behavior lies in the vector potential, which is singular upon approaching the origin. The problem can be cured by regularizing the vector potential. One possibility is to let the flux $\Phi$ have an $r$-dependence, such that $\Phi=0$ for $r<\rho$ and $\Phi=1/2$ for $r>\rho$, {\em i.e.}, the flux is not located at the origin, but on a circle of radius $\rho$. Obviously, the problem is now well-defined at the origin, and we can take the solution from the case without flux tube,
\begin{equation}
 \label{eq:mu01}
 \psi(\vr)=c \vect{e^{-i \theta} J_{1}(V_0 r)}{-i J_{0}(V_0 r)},
\end{equation}
where $c$ is a complex constant. We then match the wavefunctions from Eq. \eqref{eq:mu01} and Eq. \eqref{eq:mu0} at $r=\rho$. Upon taking $\rho\rightarrow 0$, we get $b_2=0$ as a condition for Eq. \eqref{eq:mu0}. 
The boundary condition at the origin ensures, that there is precisely one solution for zero angular momentum.

The $\mu=0$ state is also special outside the dot, where the wavefunction is proportional to $\frac{1}{\sqrt{r}}$ in both components. Thus it has the same decay as a free circular wave in two dimensions and, hence, it does not allow for the formation of a bound state. This conclusion is independent of the choice of the regularization of the wavefunction near $r=0$.

Summarizing: Without flux tube, the bound states are labeled by the angular momentum quantum number $m$, which takes half-integer values. For $|m|=1/2$ one has a ``quasi-bound state'', because the corresponding wavefunction is marginally non-normalizable. With a $\pi$ flux tube, the bound states are labeled by the kinematic angular momentum quantum number $\mu$, which takes integer values. There is no bound state for $\mu=0$.

\section{Two-terminal conductance}
\label{sec:twoterminal}

Following Refs.\ \onlinecite{Bardarson09,TitovOstrovsky,Schneider11} we now attach metallic source and drain contacts to the undoped graphene layer that surrounds the quantum dot. Schematically, this setup is shown in Fig.\ \ref{fig:geometry}. We then calculate the two-terminal conductance, where bound states of the dot show up as resonant features as a function of the gate voltage $V_0$.

The contacts are included by the addition of an additional potential $U_{\rm leads}$ with\cite{Tworzydlo06}
\begin{equation}
  U_{\rm leads} = \left\{ \begin{array}{ll} 0 & \mbox{if $-L/2 < x < L/2$}, \\
  \infty & \mbox{if $x < -L/2$ or $x > L/2$}. \end{array} \right.
\end{equation}
We apply periodic boundary conditions in the $y$ direction, with period $W$. For the vector potential $\vA$ we take a different gauge than in Sec.\ \ref{sec:boundstates},
\begin{equation}
  \vA(\vr) = \frac{h \Phi}{e} \delta(x) \ve_x \times
  \left\{ \begin{array}{ll} 0 & \mbox{if $0 < y < W/2$}, \\
  1 & \mbox{if $-W/2 < y < 0$}, \end{array} \right.
  \label{eq:Asign}
\end{equation}
where $\ve_x$ is the unit vector in the $x$ direction. With this choice of the vector potential there are two flux tubes: one, at $y=0$, located in the quantum dot, and one, at $y=W/2$, located outside the quantum dot. The second flux tube is necessary to implement the periodic boundary conditions. It does not affect the conductance resonances in the limit that the sample width $W$ is much larger than the distance $L$ between source and drain contacts.

The numerical calculation of the two-terminal conductance follows the method of Ref.\ \onlinecite{Bardarson07}. Details specific to the presence of the flux tube are discussed in the appendix. We now compare results for quantum dots with and without flux tube. We give results for a disc-shaped quantum dot, as a prototype of a quantum dot with integrable dynamics, and a stadium-shaped quantum dot, the prototype of a dot with chaotic dynamics.

\subsection{Disc-shaped dot}

\begin{figure}[t]
\includegraphics[width=2.9in]{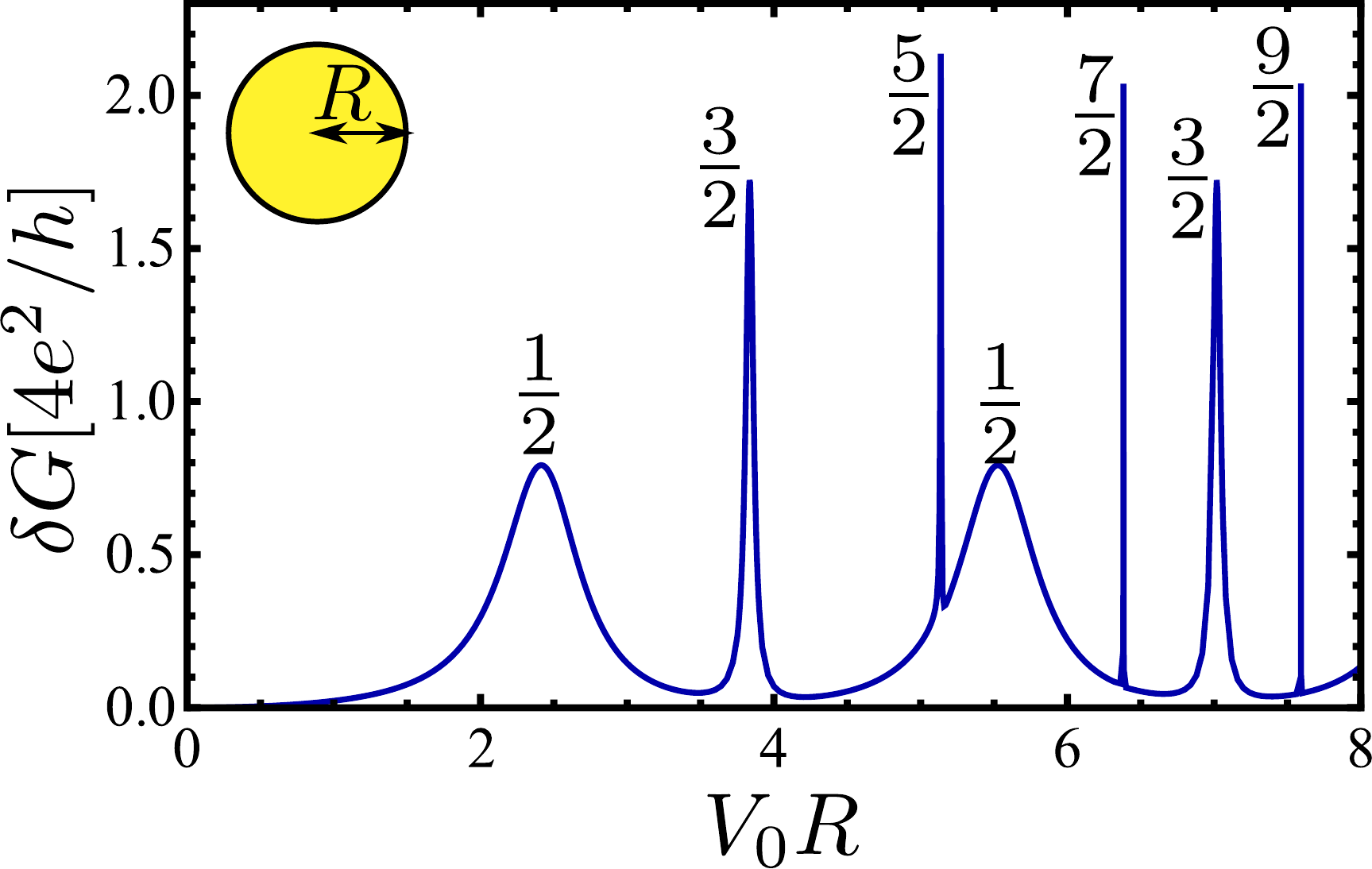}
\includegraphics[width=2.9in]{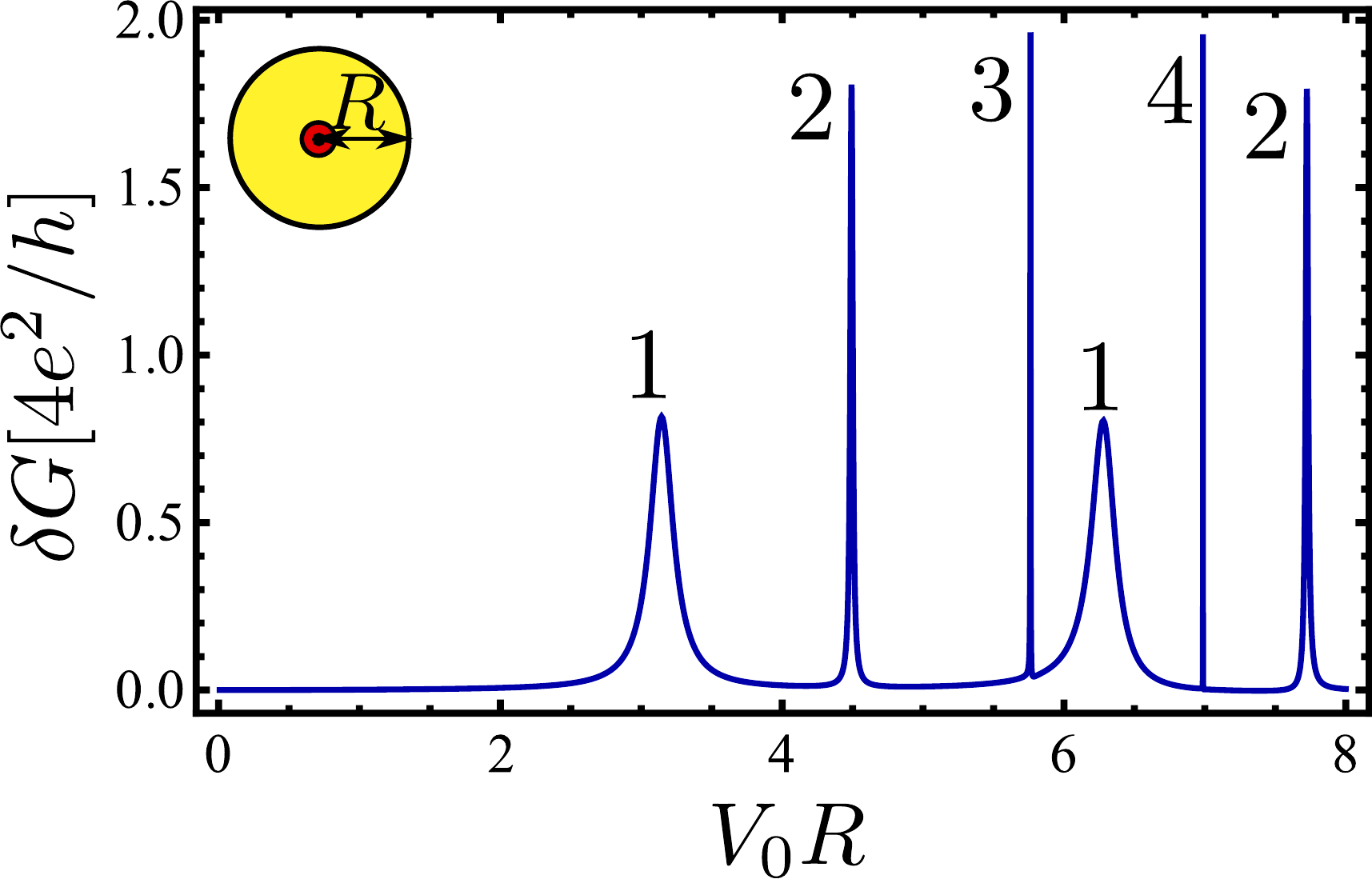}
\caption{(Color online) Two-terminal conductance of a graphene sheet containing a disc-shaped quantum dot without (top) and with (bottom) $\pi$-flux tube. Model parameters are $R/L=0.2$ and $W/L=6$. Without flux tube, resonances have definite angular momentum, with quantum number $|m|$ indicated at each resonance [data taken from Ref.\ \onlinecite{Bardarson09}]. Without flux tube, resonances are labeled by the kinematic angular momentum quantum number $|\mu|$. No resonance is found for $\mu=0$.}
\label{fig:circle}
\end{figure}

\begin{figure}[t]
\includegraphics[width=2.9in]{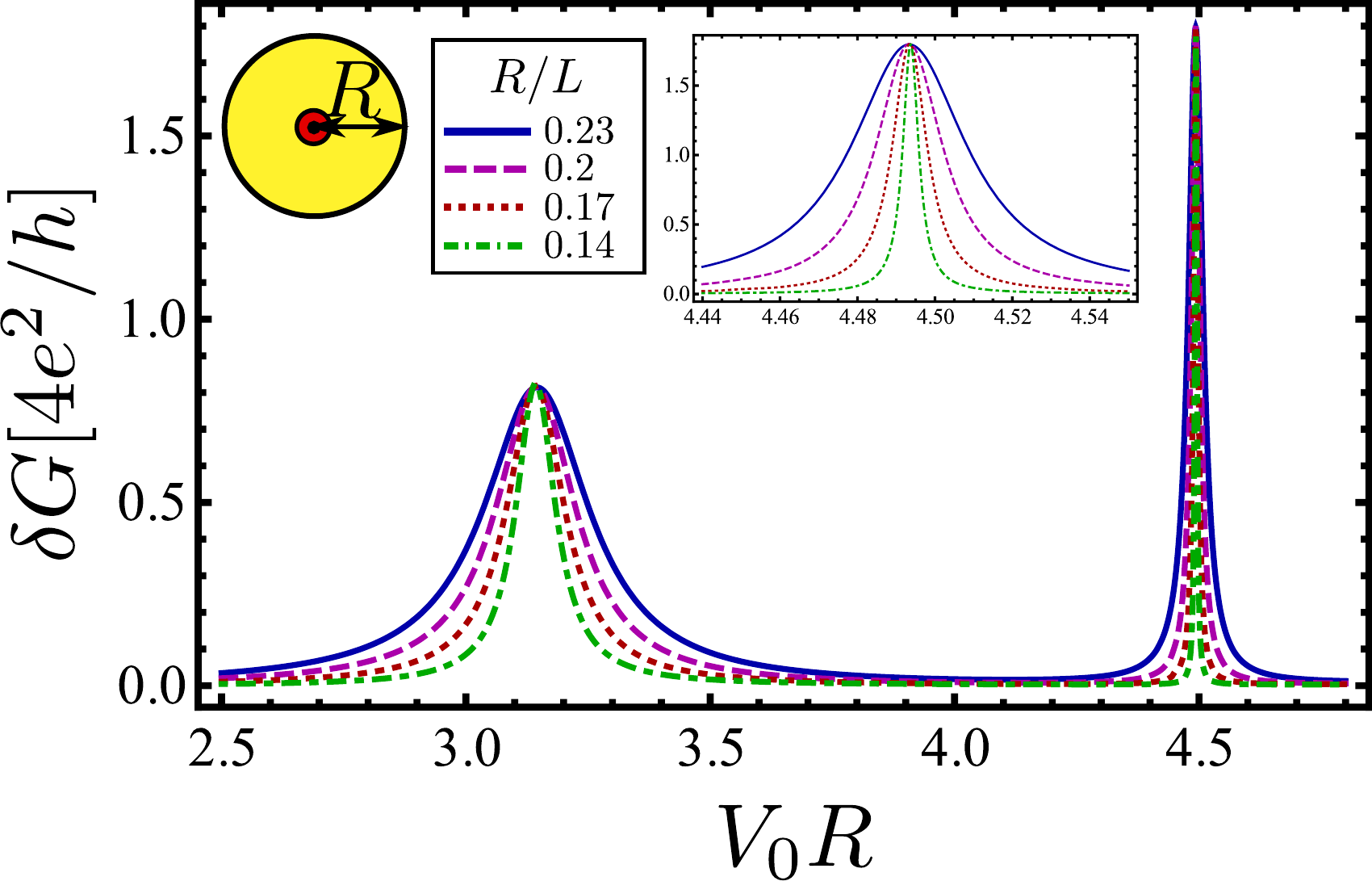}
\caption{(Color online) First two resonances for a disc-shaped quantum dot with $\pi$-flux tube, for different coupling strengths to the leads. Calculations are performed for $W/L = 8$ and various $R/L$, as indicated in the figure. The second resonance is shown enlarged in the inset.}
\label{fig:circlewithflux2}
\end{figure}

The two-terminal conductance for the case of a disc-shaped quantum dot without and with flux tube is shown in Fig.\ \ref{fig:circle}. The figure shows pronounced resonances as a function of the gate voltage $V_0$, with positions that agree with the ones calculated Sec.\ \ref{sec:boundstates}. Without flux tube, the resonances are labeled by the quantum number $|m| = 1/2$, $3/2$, $5/2$, \ldots. Their width scales $\propto (R/L)^{2|m|}$ as the coupling to the leads is decreased (data not shown), consistent with Eq.\ (\ref{eq:GammaNoFlux}) and Refs.\ \onlinecite{Bardarson09,TitovOstrovsky,Schneider11}. With flux tube, the resonances are labeled by the kinematic angular momentum quantum number $|\mu|=1$, $2$, $3$, \ldots. There are no resonances for $\mu = 0$. Upon decreasing the coupling to the leads, the resonances become narrower but retain their height, see Fig.\ \ref{fig:circlewithflux2}, and the scaling of the resonance width with the ratio $R/L$ is consistent with Eq.\ (\ref{eq:GammaFlux}) (data not shown)
.

\subsection{Stadium-shaped dot}

\begin{figure}[t]
\includegraphics[width=2.9in]{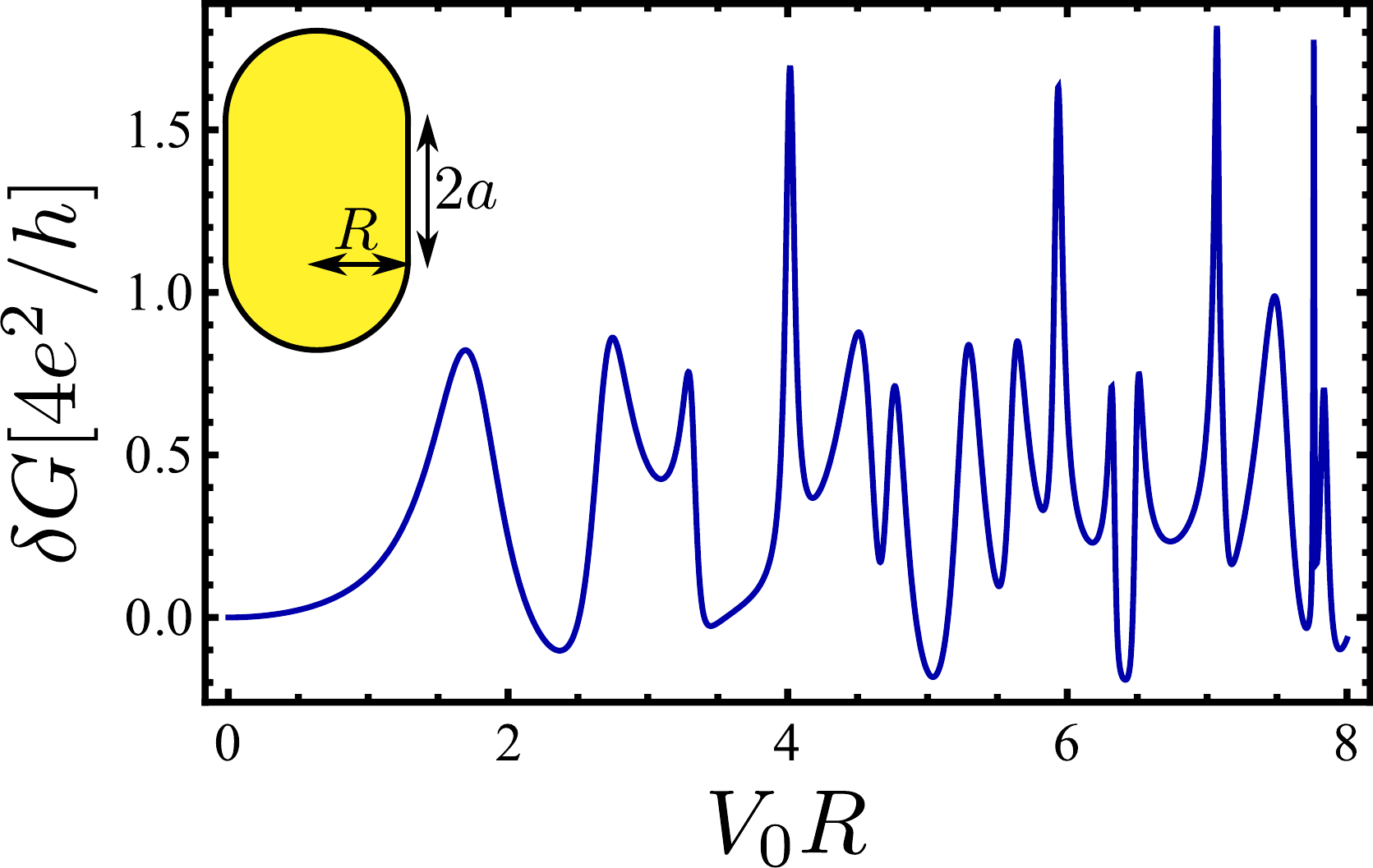}
\includegraphics[width=2.9in]{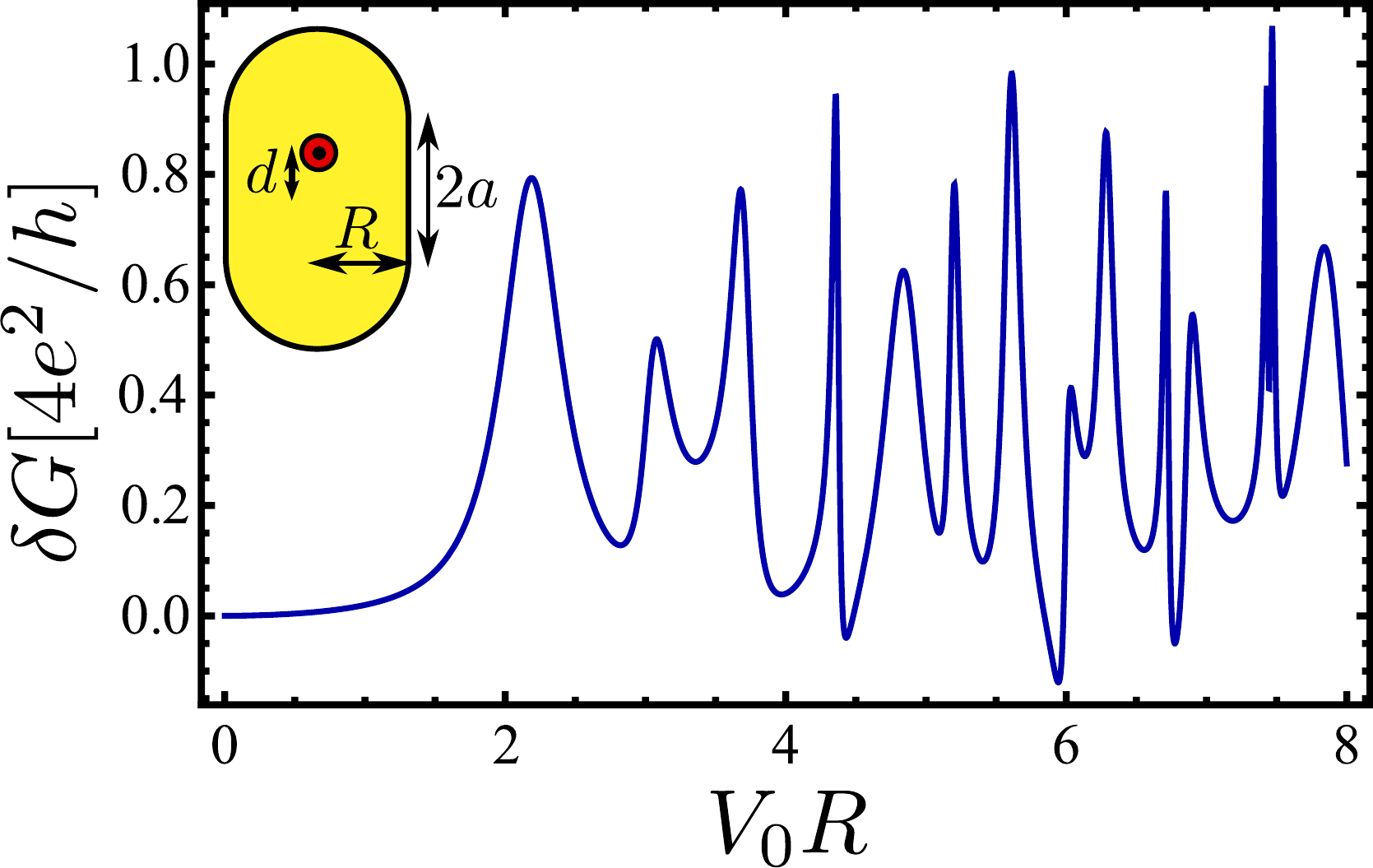}
\caption{(Color online) Two-terminal conductance of a graphene sheet containing a stadium-shaped quantum dot without (top) and with (bottom) a $\pi$-flux. Parameters for the calculation are $R/L=0.2$, $W/L=12$, $a/R=\sqrt{3}/2$, $d=2a/3$. Without flux tube, the calculation for the conductance was done with the method of Ref.\ \onlinecite{Schneider11}.}
\label{fig:stadium}
\end{figure}

\begin{figure}[t]
\includegraphics[width=2.9in]{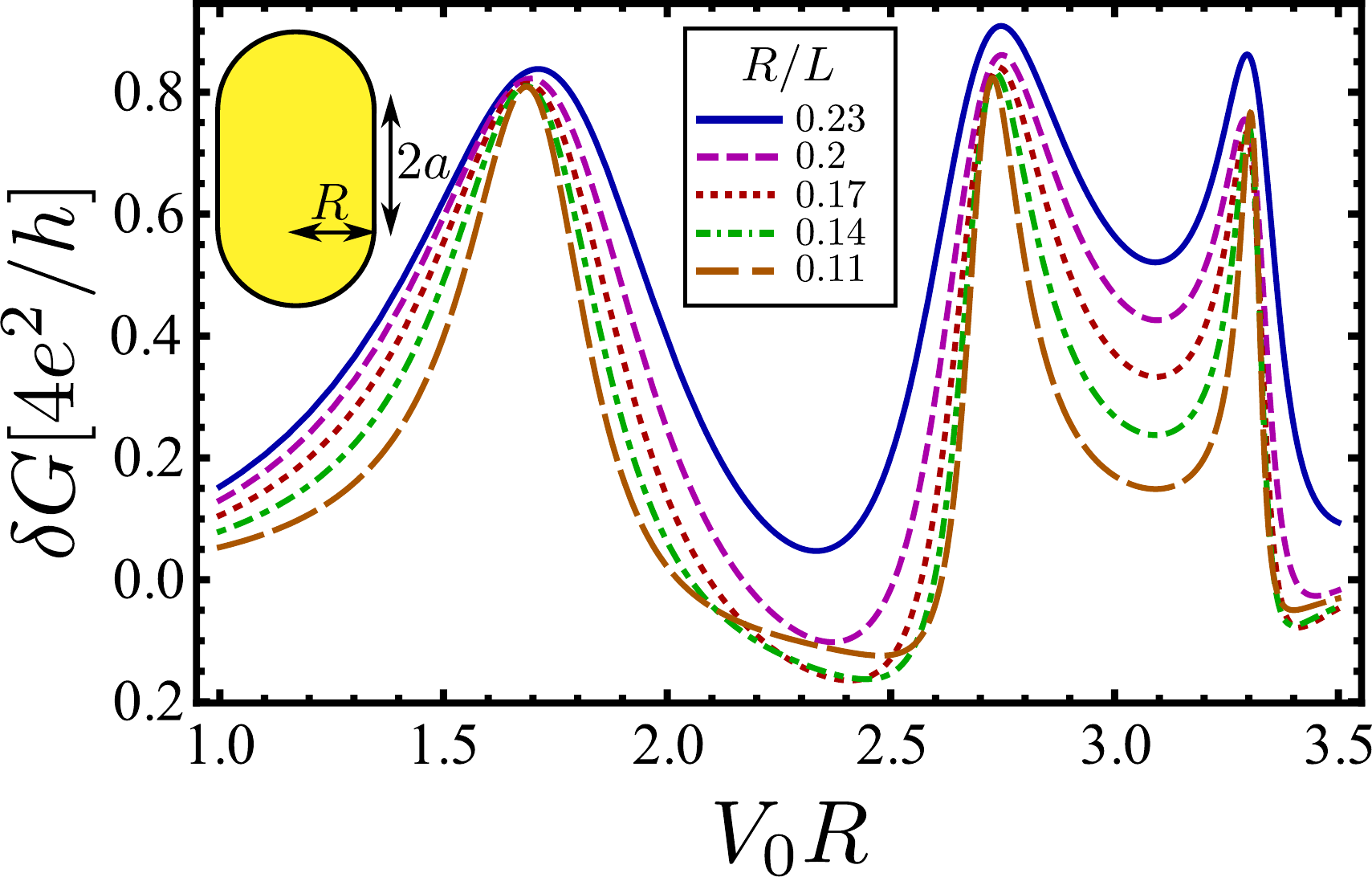}
\includegraphics[width=2.9in]{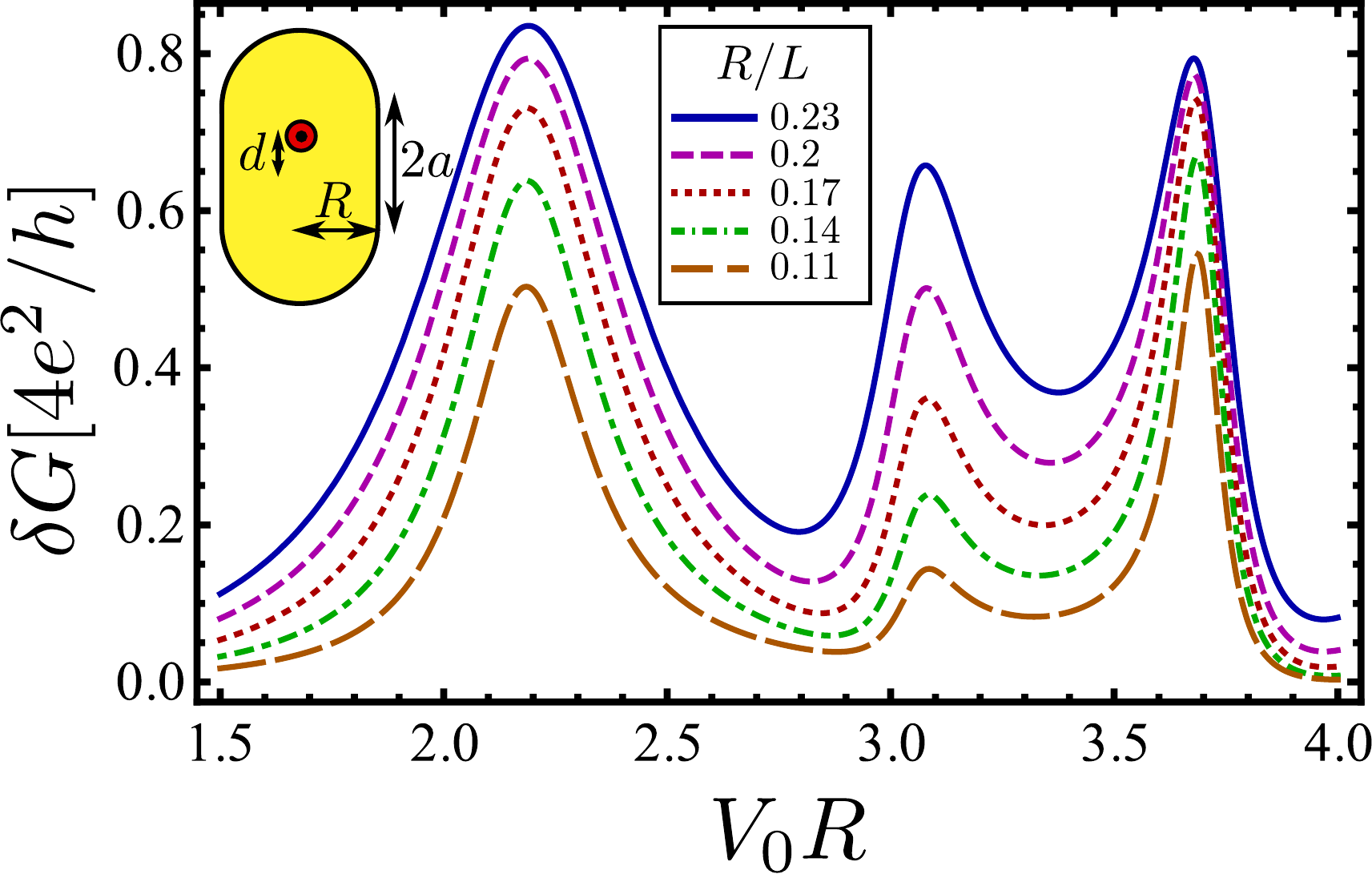}
\caption{(Color online) Behavior of the first three quasi resonances of the stadium-shaped quantum dot without (top) and with (bottom) $\pi$-flux upon changing the coupling to the leads $R/L$. The other parameters are the same as in Fig.\ \ref{fig:stadium}.}
\label{fig:stadiumRL}
\end{figure}

As a prototypical example of a chaotic quantum dot, we consider a stadium-shaped quantum dot. Here the potential $V(\vr) = - \hbar v_{\rm F} V_0$ for positions  $\vr$ inside the stadium and $V(\vr) = 0$ otherwise. Without magnetic flux, the two-terminal conductance shows resonances, which, in the limit of small $R/L$, all behave as the $|m|=1/2$-type resonances of the disc-shaped dot, {\em i.e.}, their height remains finite, whereas the resonance width scales proportional to $R/L$.\cite{Schneider11} The numerical data shown in the top panels of Figs.\ \ref{fig:stadium} and \ref{fig:stadiumRL} clearly reveal these resonances, although the asymptotic scaling of the resonance width and resonance height with $R/L$ is somewhat obscured by transient contributions for moderate $R/L$ that originate from higher-angular-momentum contributions to the resonances.\cite{Schneider11}

The conductance trace for a stadium-shaped quantum dot with a flux tube carrying half a flux quantum is shown in the bottom panels of Figs.\ \ref{fig:stadium} and \ref{fig:stadiumRL}. In order to break inversion symmetry, the stadium is placed asymmetrically with respect to the flux tube, see the inset of Fig.\ \ref{fig:stadium}. The differences with the case of the disc-shaped quantum dot and with the case without a flux tube are significant. We find that the conductance depends on the gate voltage $V_0$ for finite $R/L$, but the widths of the ``resonances'' is independent of the coupling to the leads, which is set by the ratio $R/L$, whereas the height decreases upon decreasing $R/L$. This agrees with the expectation that, since all states in the stadium have a $\mu=0$ component, a stadium dot should not support any (quasi)bound states. While for intermediate values of $R/L$ contributions from higher angular momentum channels still give rise to broad ``quasi-resonances'', in the limit $R/L \to 0$, only the 
$\mu=0$ channel is relevant, and the conductance becomes featureless as a function of $V_0$.

We remark that, if the flux tube would be placed exactly in the middle of the stadium, inversion symmetry would split the resonances into two groups, resulting from even and odd $\mu$. The ``even'' resonances have a finite $\mu=0$ component and disappear upon taking the limit $R/L \to 0$. The ``odd'' resonances survive in this limit, with a finite resonance height and a resonance width $\Gamma \propto (R/L)^2$ (data not shown).

\section{Conclusion}
\label{sec:conclusion}

In this article we investigated the observation of Refs.\ \onlinecite{Bardarson09,TitovOstrovsky,Schneider11}, that the two-terminal conductance of a generic gate-defined graphene quantum dot shows resonances in the limit of a weak coupling to the leads, in spite of the naive expectation that electrons can not be confined in such a quantum dot because of Klein tunneling. We attribute this observation to the Berry phase in graphene, which quantizes angular momenta to half-integer values. With half-integer angular momenta, strict perpendicular incidence --- the condition for Klein tunneling with unit probability --- does not occur. As a consequence, conductance resonances exist in both integrable and chaotic geometries. The only difference between the two cases is a quantitative one: it concerns the scaling of the resonance widths with the coupling to the leads.\cite{Schneider11}

The Berry phase can be cancelled against an Aharonov-Bohm phase, when a flux tube containing half a flux quantum is introduced to the system. With a magnetic flux tube, we showed that the relevant angular momentum, the kinematical angular momentum, is quantized to integer values. In this case a state with zero angular momentum is possible. Such a state can not form a bound state or give rise to a conductance resonance. We showed this by an explicit calculation for the disc-shaped quantum dot in Sec.\ \ref{sec:boundstates}, and using numerical calculations for disc-shaped and stadium-shaped quantum dots in Sec. \ref{sec:twoterminal}. Once the Aharonov-Bohm phase from the $\pi$-flux tube cancels the Berry phase, the results of the full quantum theory are consistent with the simple classical expectations. With a $\pi$ flux, there is a stark qualitative difference between conductance resonance for integrable and non-integrable quantum dots: Whereas sharp conductance resonances for the case of an integrable 
quantum dot continue to exist, in the limit of weak lead-dot coupling the conductance becomes featureless for a generic non-integrable quantum dot.

In closing, we make two remarks concerning the possible realization of the scenario we investigated here. First, a flux tube for graphene not necessarily has to be created by a real magnetic field, but it can also be engineered via strain as a pseudo-magnetic field.\cite{Fogler,Levy} The pseudo-magnetic field would have opposite signs for the two valleys, which is of no consequence for our conclusions, because the two valleys are decoupled for the smooth confining potentials we consider here. Second, the application of a well-defined Aharonov-Bohm phase is more controlled in ring-shaped structures.\cite{yau2002,qu2011} In this case, we expect that the our main finding, the strong qualitative difference between integrable and non-integrable geometries in the case of a $\pi$ flux, persists.

\acknowledgments
We gratefully acknowledge discussions with Jens Bardarson, Igor Gornyi, Pavel Ostrovsky, Brian Tarasinski, and Silvia Viola Kusminskiy. This work is supported by the Alexander von Humboldt Foundation in the framework of the Alexander von Humboldt Professorship, endowed by the Federal Ministry of Education and Research and by the German Research Foundation (DFG) in the framework of the Priority Program 1459 ``Graphene''.

\appendix

\section{Numerical approach}

The numerical approach follows Ref.\ \onlinecite{Bardarson07}. The potential $V(\vr)$ is replaced by a potential $\sum_{n} V_n(y) \delta(x-x_n)$ that is nonzero at $N$ discrete values $-L/2 = x_0 < x_1 < x_2 < \ldots < x_{N-1} < x_N = L/2$ of the $x$ coordinate, with 
$$
  V_n(y) = \int_{(x_{n-1}+x_{n})/2}^{(x_{n}+x_{n+1})/2} dx V(x,y),\ \
  n=1,2,\ldots,N-1.
$$%
Between the discrete points the wavefunction is solved from the free Dirac equation. This solution takes its simplest form after Fourier transform with respect to the transverse coordinate $y$, because the free Dirac equation does not couple different transverse modes. These solutions are then matched by applying the appropriate boundary conditions at the discrete points $x=x_j$, $j=1,2,\ldots,N$. A numerically stable method to implement this program is to express both the solution of the free Dirac equation and the matching conditions at the discrete points $x=x_j$ in terms of scattering matrices. The scattering matrix of the entire sample is then obtained from convolution of the scattering matrices of the $2 N - 1$ individual components. The result of the calculation is the transmission matrix $t$, which is related to the two-terminal conductance via the Landauer formula,
\begin{equation}
  G = \frac{4 e^2}{h} \mbox{tr}\, t t^{\dagger},
\end{equation}
where the trace is taken of the transverse Fourier modes. The number of ``slices'' $N$ and the number of transverse modes $M$ must be chosen large enough, that the conductance $G$ no longer depends on $N$ and $M$.

The vector potential (\ref{eq:Asign}) corresponds to the boundary condition 
\begin{equation}
  \lim_{x \uparrow 0} \psi(x,y) = - \lim_{x \downarrow 0} \psi(x,y)\ \
  \mbox{for $-W/2 < y < 0$},
\end{equation}
whereas $\psi(x,y)$ is continuous at $x=0$ for $0 < y < W/2$. In the approach described above, this boundary condition is expressed in terms of a scattering matrix relating incoming and outgoing waves at $x \uparrow 0$ and $x \downarrow 0$. This scattering matrix has no reflective part, whereas the transmission matrix is
\begin{equation}
  t_{mn} = \left\{ \begin{array}{ll} 0 & \mbox{if $m-n$ even}, \\
  -4 i/[(k_m-k_n)W] & \mbox{if $m-n$ odd},
  \end{array} \right. \label{eq:tmatrix}
\end{equation}
where the integers $m$ and $n$ label the transverse modes and $k_n = 2 \pi n/W$. This transmission matrix has the special properties that $t = t^{\dagger}$ and $t^2 = 1$.

In order to ensure numerical stability, unitarity must be preserved while restricting to a finite number of transverse modes $M$. For the transmission matrix (\ref{eq:tmatrix}) this can be achieved using the following trick: One first builds the hermitian matrix $h = i (e^{i \phi} - t)/(t + e^{i \phi}) = \cot \phi - t/\sin \phi$ out of the transmission matrix, where $\phi$ is a phase that can be chosen arbitrarily, then truncates $h$, which can be done straightforwardly without compromising hermiticity, and then uses the inverse relation $t = e^{i \phi} (1+ ih)/(1 - ih)$ to obtain a finite-dimensional transmission matrix. In our numerical calculation we set $\phi=\pi/2$. We verified that the elements of the resulting finite-dimensional transmission matrix approach the elements of the exact transmission matrix (\ref{eq:tmatrix}) in the limit that the number of transverse mode $M \to \infty$.

\end{document}